\documentclass[10pt,twocolumn,aps,prl,superscriptaddress,reprint]{revtex4-2}
\usepackage{setspace}
\usepackage{graphicx}% Include figure files
\usepackage{float}
\usepackage{ulem}
\usepackage{dcolumn}% Align table columns on decimal point
\usepackage{hyperref}
\usepackage{multirow}
\usepackage{xcolor}
\usepackage{color}
\usepackage{soul,xcolor}
\usepackage{lineno}
\usepackage{amsmath}

\begin{document}

\title{Metal-to-superconductor Transition Induced by Lithium Adsorption on Monolayer 1$T$-Nb$_2$C}% Force line breaks with \\

\author{Lingxiao Xiong}
 \affiliation{College of Physics \& Optoelectronic Engineering, Department of Physics, Jinan University, Guangzhou 510632, China}

\author{Yuhui Yan}%
 \affiliation{College of Physics \& Optoelectronic Engineering, Department of Physics, Jinan University, Guangzhou 510632, China}

\author{Feipeng Zheng}%
 \email{fpzheng\_phy@email.jnu.edu.cn}
 \thanks{corresponding author}
 \affiliation{College of Physics \& Optoelectronic Engineering, Department of Physics, Jinan University, Guangzhou 510632, China}

\date{\today}

\begin{abstract}
    Recently, two-dimensional Nb$_2$C has garnered increasing attention, both experimentally and theoretically, due to its functional-group-dependent superconductivity.
    In contrast to the halogen and chalcogen additives that have been the main focus of previous studies, we study the effect of lithium adsorption, which can also be incorporated during the synthesis of Nb$_2$C.
    Our computational analysis reveals a metal-to-superconductor transition in monolayer Nb$_2$C with a critical temperature ($T_{\mathrm{c}}$) of 22.1 K and a strong anisotropic superconducting gap distribution following the adsorption of lithium atoms.
    This emergent superconductivity is attributed to the increased electronic states at the Fermi energy, resulting from the contribution of Nb-$d$ orbitals and electron gas states induced by the low electronegativity of lithium.
    Furthermore, the application of tensile strain raises the $T_{\mathrm{c}}$ to 24 K, which is higher than that of most functional-group-modified Nb$_2$C systems.
    Our work deepens the understanding of electron-phonon coupling in layered Nb$_2$C, and provides new insights into achieving high critical temperature superconductivity with a strong anisotropic superconducting gap distribution in this system.
\end{abstract}

\maketitle

\section{introduction}

Two-dimensional materials have garnered increasing attention due to emergent ordering that is absent in bulk materials \cite{lu2015evidence}, structural or electronic phase transitions when transitioning from bulk to monolayer forms \cite{qian2014quantum,zheng2016quantum,zhang2023emergent,ellis2011indirect}, and their easily tunable properties.
Recently, MXenes---a novel class  of two-dimensional transition metal carbides or nitrides---are of particular interest for their experimental feasibility for synthesis \cite{kamysbayev2020covalent,li2024mxenes,xu2015large}, highly tunable properties \cite{rong2024elastic,hong2024multigap,zhang2024tailoring,wang2024magnetically}, and potential applications in superconductivity \cite{zhao2023nonlinear,yin2024origin} and energy storage \cite{shi2023redox,saraf2024enhancing,mahajan20241d,lu2024theoretical}.
Particularly, layered 1$T$-Nb$_2$C (henceforth Nb$_2$C) systems have received considerable attention,  due to their functional-groups-dependent superconductivity, despite the pristine form being non-superconducting \cite{kamysbayev2020covalent,wang2022role,jin2023structural,li2023robust}.
Experimentally, the superconductivity was not detected in bulk Nb$_2$C down to 1.8 K. 
However, it was observed in Nb$_2$CS$_2$, Nb$_2$CSe$_2$ and Nb$_2$C(NH) with  $T_{\mathrm{c}}$s below 7.1 K \cite{kamysbayev2020covalent}.
This finding was further corroborated by experimental and first-principles studies \cite{wang2022role,jin2023structural}. 
In contrast, Nb$_2$CO$_x$ and Nb$_2$CF$_x$ did not show superconductivity \cite{kamysbayev2020covalent,jin2023structural}. 
Theoretical calculations indicate that monolayer Nb$_2$C exhibits similar superconducting behavior to its bulk form \cite{bekaert2020first,yang2020electronic,wang2022straintronic,bekaert2022enhancing,sevik2023superconductivity,zhang2023effect}.
Although pristine monolayer Nb$_2$C is non-superconducting \cite{bekaert2020first}, the adsorption of halogen atoms (Cl \cite{sevik2023superconductivity,zhang2023effect}, Br \cite{yang2020electronic,zhang2023effect}), chalcogen atoms (O \cite{yang2020electronic,wang2022straintronic,zhang2023effect}, S \cite{wang2022straintronic,sevik2023superconductivity,zhang2023effect}, Se \cite{wang2022straintronic,sevik2023superconductivity,zhang2023effect}, Te \cite{wang2022straintronic}, SO \cite{zhang2023effect}, SSe \cite{zhang2023effect}), and H atoms \cite{yang2020electronic,bekaert2022enhancing} can induce the superconductivity with $T_{\mathrm{c}}$s ranging from 1 K to 38 K.
These studies conclude that the superconductivity in monolayer Nb$_2$C can be significantly enhanced by the adsorption of halogen, chalcogen, and hydrogen atoms.
Nonetheless, some issues still need to be resolved.

Firstly, previous studies primarily focused on the adsorption of halogen, chalcogen and hydrogen, which are easily introduced during the synthesis of Nb$_2$C \cite{lim2022fundamentals}. 
However, alkali metal atoms like lithium (Li) and sodium (Na) can also be incorporated during MXene synthesis \cite{liu2017preparation,lipatov2016effect}, and their deposition can be achieved experimentally \cite{ludbrook2015evidence,kondekar2017situ,li2020reduced,yang2022high,lee2024interaction}.
Furthermore, as the crystal structure of Nb$_2$C is terminated by metallic Nb atoms with smaller electronegativity than the halogen, chalcogen and H atoms, this raises a question of whether alkali metal atoms, which have much lower electronegativity, can be energetically favorable for adsorption on the Nb$_2$C surface and how their adsorption affects the material’s superconductivity compared to halogen, chalcogen, and hydrogen atom adsorption.
Finally, a recent study experimentally highlighted the anisotropic superconductivity in Nb$_2$C decorated with functional groups \cite{xu2024anisotropic}, making it crucial to explore these anisotropic properties, which have largely been overlooked in previous studies.

In this work, we employed first-principles calculations to investigate the electronic structure, electron-phonon coupling, and anisotropic superconductivity of Li-adsorbed $\mathrm{Nb_2C}$. 
We found that the Li adsorption is energetically favorable to form the crystal structure with a chemical formula of  $\mathrm{Nb_2CLi_2}$.
The adsorption of Li atoms significantly increases the density of states at the Fermi level ($N(0)$), arising from electron doping due to the introduction of Li, and the emergent delocalized electronic gas states resulting from the redistribution of Li's valence electrons.
These effects substantially enhance electron-phonon coupling (EPC), transforming the system from a conventional metal into a single-gap superconductor with a critical temperature ($T_{\mathrm{c}}$) of 22 K and strong anisotropic superconducting properties characterized by an extended gap energy range.
The application of tensile strain further increases the $T_{\mathrm{c}}$ to 24 K.

\section{COMPUTATIONAL METHOD}

Density-functional theory and density-functional perturbation theory calculations were performed using the PBE exchange-correlation functional \cite{perdew1996generalized}, to scrutinize the crystal structures, electronic structures, and electron-phonon coupling (EPC) of monolayer Nb$_2$C before and after alkali-metal adsoprtions\cite{kresse1996efficient,giannozzi2009quantum,giannozzi2020quantum,ponce2016epw}.
The norm-conserving pseudopotential of FHI98 \cite{fuchs1999ab} was used to describe the interaction between valance and core electrons.
These calculations employ a combination of the QUANTUM ESPRESSO (QE) \cite{giannozzi2009quantum,giannozzi2020quantum} and Vienna Ab initio Simulation Package (VASP) \cite{kresse1996efficient}.
VASP is used for calculating the formation energy and $ab$ $initio$ molecular dynamics (AIMD) simulations \cite{allen2017computer}, while the remaining calculations are performed using QE.
A vacuum layer with a thickness of 12.5~\AA\ was introduced, which is enough to eliminate interactions between periodic images as shown in  Sec. S5 \cite{SM}.

The crystal structures were visualized using {\scriptsize VESTA} package \cite{momma2011vesta}.
The Kohn-Shame valence states were expanded using plane waves with an energy cutoff of 80 Ry.
An 18$\times$18$\times$1 $\boldsymbol{k}$ mesh and a 6$\times$6$\times$1 $\boldsymbol{q}$ mesh were used to calculate the ground states of charge densities and phonons for the monolayer systems, whereupon the EPC matrix elements, $g_{mn,\nu}(\boldsymbol{k},\boldsymbol{q})$, were calculated \cite{ponce2016epw}.
The matrix element quantify the scattering amplitude between the electronic states with a wave vector $\boldsymbol{k}$, a band index $m$ [denoted as ($\boldsymbol{k},m)$], and ($\boldsymbol{k}$+$\boldsymbol{q}$, n)  through a phonon mode with a branch $\nu$ and a wave vector $\boldsymbol{q}$.
The above quantities were further interpolated \cite{mostofi2008wannier90} to a $120 \times 120 \times 1$ $\boldsymbol{k}$ grid and a $60 \times 60 \times 1$ $\boldsymbol{q}$ grid, whereby the Eliashberg function $\alpha^2F(\omega)$ was calculated.
To ensure the accuracy of our electronic structure calculations, we constructed Wannier functions using a total of twelve orbitals, including all the Nb-d and Li-s orbitals.
We compared the band structures obtained from density functional theory (DFT) and those from the Wannier functions, which showed good consistency, as demonstrated in Sec.~S5 \cite{SM}.
The Eliashberg function is defined as
\linenomath
$$
\alpha^2F(\omega)=\frac{1}{2}\sum_{\nu}\int_{\mathrm{BZ}}\frac{\mathrm{d}\boldsymbol{q}}{\Omega_{\mathrm{BZ}}}\omega_{\boldsymbol{q}\nu}\lambda_{\boldsymbol{q}\nu}\delta\left(\omega-\omega_{\boldsymbol{q}\nu}\right),
$$
\endlinenomath
where $\lambda_{\boldsymbol{q}\nu}$ is a phonon-momentum-resolved EPC constant, $\Omega_{\mathrm{BZ}}$ the volume of the first Brillouin zone. 
The Dirac delta function $\delta\left(\omega-\omega_{\boldsymbol{q}\nu}\right)$ is approximated by a Gaussian function with a broadening of 0.5 meV.
After testing, these parameters are found to be sufficient to obtain converged and reliable results, as shown in Sec.~S5 \cite{SM}.
The atomic-vibration-resolved Eliashberg function $\alpha^2F(\omega,j,\hat{n})$ is defined as 
\linenomath
$$
  \alpha^{2}F(\omega,j,\hat{n})=\frac{1}{2}\sum_{v}\int_{\mathrm{BZ}}\frac{d\mathbf{q}}{\Omega_{\mathrm{BZ}}}\omega_{\mathbf{q}\nu}\lambda_{\mathbf{q}\nu}\delta(\omega-\omega_{\mathbf{q}\nu})\big|\hat{n}\cdot e_{\mathbf{q},\nu}^{j}\big|^{2},
$$
\endlinenomath
where $e^{j}_{q,\nu}$ is the component of atom $j$ in the eigenvector of the dynamic matrix corresponding to phonon momentum $\mathbf{q}$ and modes $\nu$. 
The unit projection direction vector $\hat{n}$ is selected along either the in-plane (xy) or out-of-plane (z) directions.
The total EPC $\lambda$ can be calculated by the intergration of Eliashberg function $\alpha^2F(\omega)$\cite{allen1975transition}
$$
\lambda(\omega)=\sum_{\boldsymbol{q}\nu}\lambda_{\boldsymbol{q}\nu}=2\int_0^\omega\frac{\alpha^2F(\omega)}{\omega}\mathrm{d}\omega.
$$
The isotropic superconducting $T_{\mathrm{c}}$ is determined by the calculated  $\lambda$ via the full Allen-Dynes formula
$$
T_{\mathrm{c}}=f_1f_2\frac{\omega_{\log}}{1.2}\mathrm{exp}\biggl[-\frac{1.04(1+\lambda)}{\lambda-\mu^*(1+0.62\lambda)}\biggr],
$$
with 
$$
\begin{aligned}f_{1}f_{2}&=\sqrt[3]{1+\left(\frac{\lambda}{2.46(1+3.8\mu^{*})}\right)^{\frac{3}{2}}}\\&\times (1-\frac{\lambda^{2}(1-\omega_{2}/\omega_{\log})}{\lambda^{2}+3.312(1+6.3\mu^{*})^{2}}),\end{aligned}
$$
where $\mu^*$ is the effective screened Coulomb repulsion constant, $\omega_{\log}$ is the logarithmic average frequency,
$$
\omega_{\log}=\exp\biggl[\frac{2}{\lambda}\int_{0}^{\infty}\frac{d\omega}{\omega}\alpha^{2}F(\omega)\log\omega\biggr],
$$
and $\omega_2$ is the mean-square frequancy,
$$
\omega_2=\sqrt{\frac{1}{\lambda}\int_0^{\omega_{\max}}\left[\frac{2\alpha^2F(\omega)}{\omega}\right]\omega^2d\omega}.
$$
Similar to previous studies\mbox{\cite{sevik2023superconductivity,yang2020electronic}}, the $\mu^* = 0.1$ is used in this work.
For completeness, the dependence of $T_c$ on different values of $\mu^{*}$ within the range of 0.1 to 0.2 is also shown in Sec.~S5 \cite{SM} for reference.
The electron-momentum-resolved superconducting gaps on the Fermi surface at temperature $T$, denoted as $\Delta(\boldsymbol{k},T)$, were determined by solving the anisotropic Midgal-Eliashberg equations on an imaginary axis, and then analytically continued to the real axis using Padé functions \cite{margine2013anisotropic}.
In solving the equations, the Kohn-Sham states within 200 meV around the Fermi level are included, and the Matsubara frequencies are cut off at 0.55 eV, which are sufficient to describe the EPC in this system.

\section{results and discussions}

\subsection{crystal structure}
 
\begin{figure}[h]
    \includegraphics{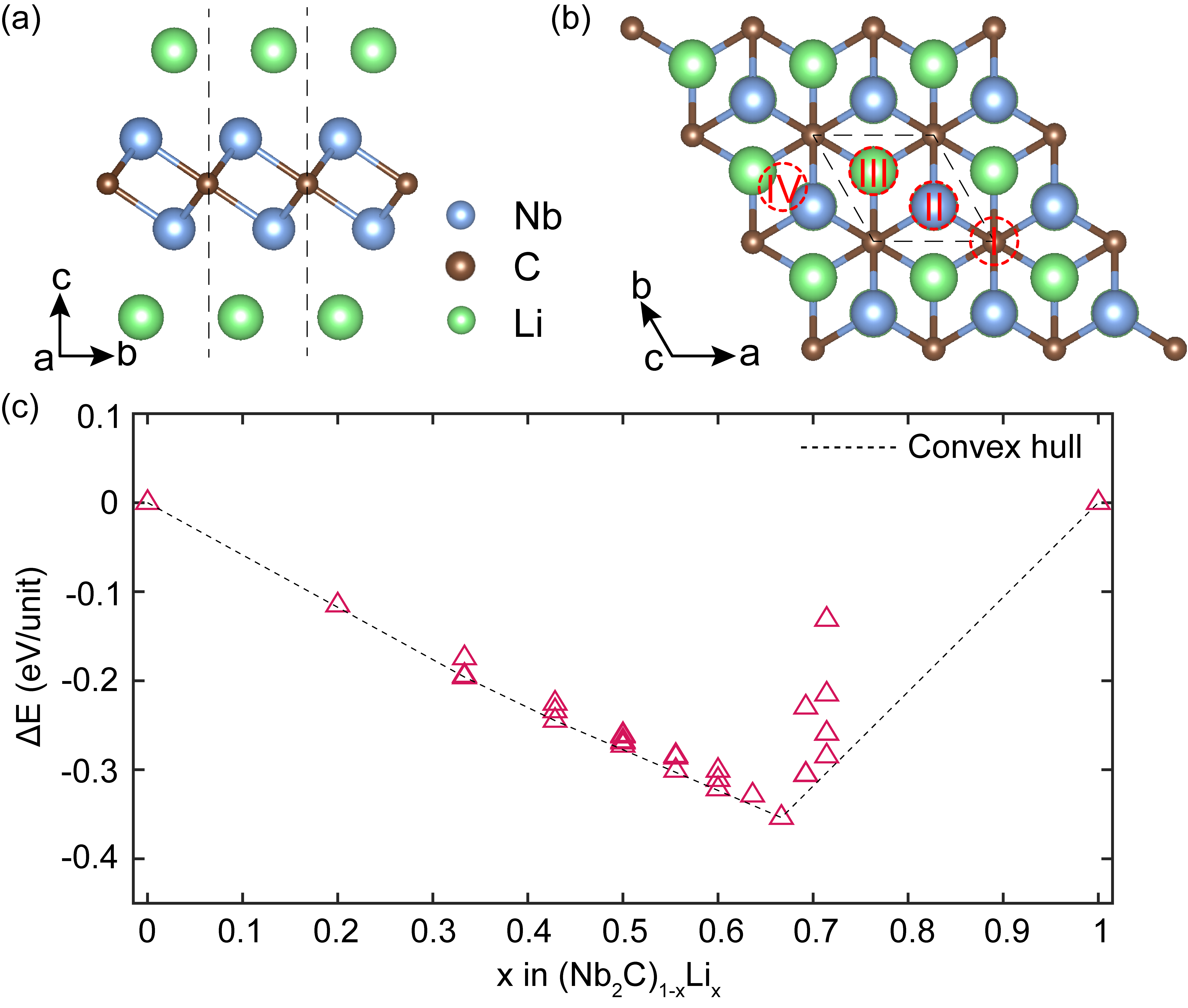}
    \caption{\label{fig1} 
    Side (a) and top views (b) of the crystal structure of $\mathrm{Nb_2CLi_2}$, corresponding to a Li concentration of $x  = 0.667$.
    The unit cell is indicated by the black dotted lines. 
    The dashed circles in panel (b) represent four kinds of potential adsorption sites. 
    (c) Formation energies of Li-adsorbed monolayer $\mathrm{Nb_2C}$ on type-III sites as a function of $x$, calculated using $2 \times 2$ supercells. 
    }
\end{figure}

Monolayer 1$T$-Nb$_2$C consists of Nb-C-Nb atomic layers, in which each  C atom is octahedrally coordinated by six Nb atoms.
Its optimized lattice constant $a=3.132$ \AA, is very close to the experimentally measured values  \cite{kamysbayev2020covalent,zaheer2022nickel}. 
When Li atoms are deposited onto the surface of pristine monolayer Nb$_2$C, there are four high-symmetry sites available for adsorption: (I) overhead C atoms, (II) overhead Nb atoms on the same side, (III) overhead  Nb atoms on the opposite side, and (IV) overhead the midpoint between two Nb atoms on different sides, as illustrated by red dashed circles in Fig.~\ref{fig1}(b).
The calculated formation energies for Li adsorptions on the aforementioned sites are -1.054, -1.044,-1.057,-1.056 eV/Nb$_{2}$C, respectively, indicating that the type-III site is the most energetically favorable position for the adsorption.
It is noteworthy that Li atoms initially placed at the type-IV sites will migrate to the type-III sites after structural optimization, which accounts for the similar formation energies listed above.
To further evaluate the thermodynamic stability of the Li-adsorbed monolayer Nb$_2$C at the type-III sites, we calculated the formation energies as a function of Li concentrations using $2\times 2$ supercells, in which there are  8 type-III sites on two sides of the monolayer.
The chemical formula of the  Li-adsorbed monolayer Nb$_2$C in the supercell can be written as (Nb$_2$C)$_{m}$Li$_{n}$, where $n$ is the number of adsorbed Li atoms, and $m = 4$ is the number of Nb$_2$C units in the supercell. 
Alternatively, it can also be written as (Nb$_2$C)$_{1-x}$Li$_{x}$, where $x = n/(m+n)$ represents the concentration of the Li atoms. 
The formation energies were calculated using the formula $\Delta E(x) = E[(\mathrm{Nb}_2\mathrm{C})_{1-x}\mathrm{Li}_{x}] -  (1-x)E(\mathrm{Nb}_{2}\mathrm{C}) - x E(\mathrm{Li})$, where $E[(\mathrm{Nb}_2\mathrm{C})_{1-x}\mathrm{Li}_{x}]$,  $E(\mathrm{Nb}_{2}\mathrm{C})$ and $E(\mathrm{Li})$ represent the total energies of (Nb$_2$C)$_{1-x}$Li$_{x}$, pristine monolayer $\mathrm{Nb}_{2}\mathrm{C}$, and Lithium crystalline in body-center cubic per unit, respectively.
It is seen from Fig.~\ref{fig1}(c) that the $\Delta E(x)$ decreases continuously with increasing $x$, until $x = 0.667$.
When more Li atoms are deposited ($x > 0.667$), the $\Delta E(x)$ starts to rise.
Consequently, the $\Delta E(x)$ exhibits its minimum value and resides on the convex hull at $x = 0.667$, as shown in Fig.~\ref{fig1}(c), indicating the dynamical stability of the corresponding structure.
Other compositions on the convex hull exhibit relatively lower $N(0)$ (see Sec.~S7 \cite{SM} for details). 
Therefore, we will focus the discussion on Nb$_2$CLi$_2$.
Further details can be found in Sec.~S7 \cite{SM}.
The side  and top views of this crystal structure are shown in Figs.~\ref{fig1}(a) and ~\ref{fig1}(b), respectively.
It is clear that all the type-III sites are occupied, leading to a chemical formula of Nb$_2$CLi$_2$.
The calculated $\omega_{\boldsymbol{q}\nu}$ shown in Fig.~\ref{fig3}(a) provides further evidence of the dynamical stability of the Nb$_2$CLi$_2$.
The results of the AIMD simulation demonstrate that the crystal structure of Nb$_2$CLi$_2$ exhibits thermodynamic stability at 300 K (refer to Sec.~S8 \cite{SM} for details).

\subsection{Electronic structure}

\begin{figure*}[htb]
    \includegraphics{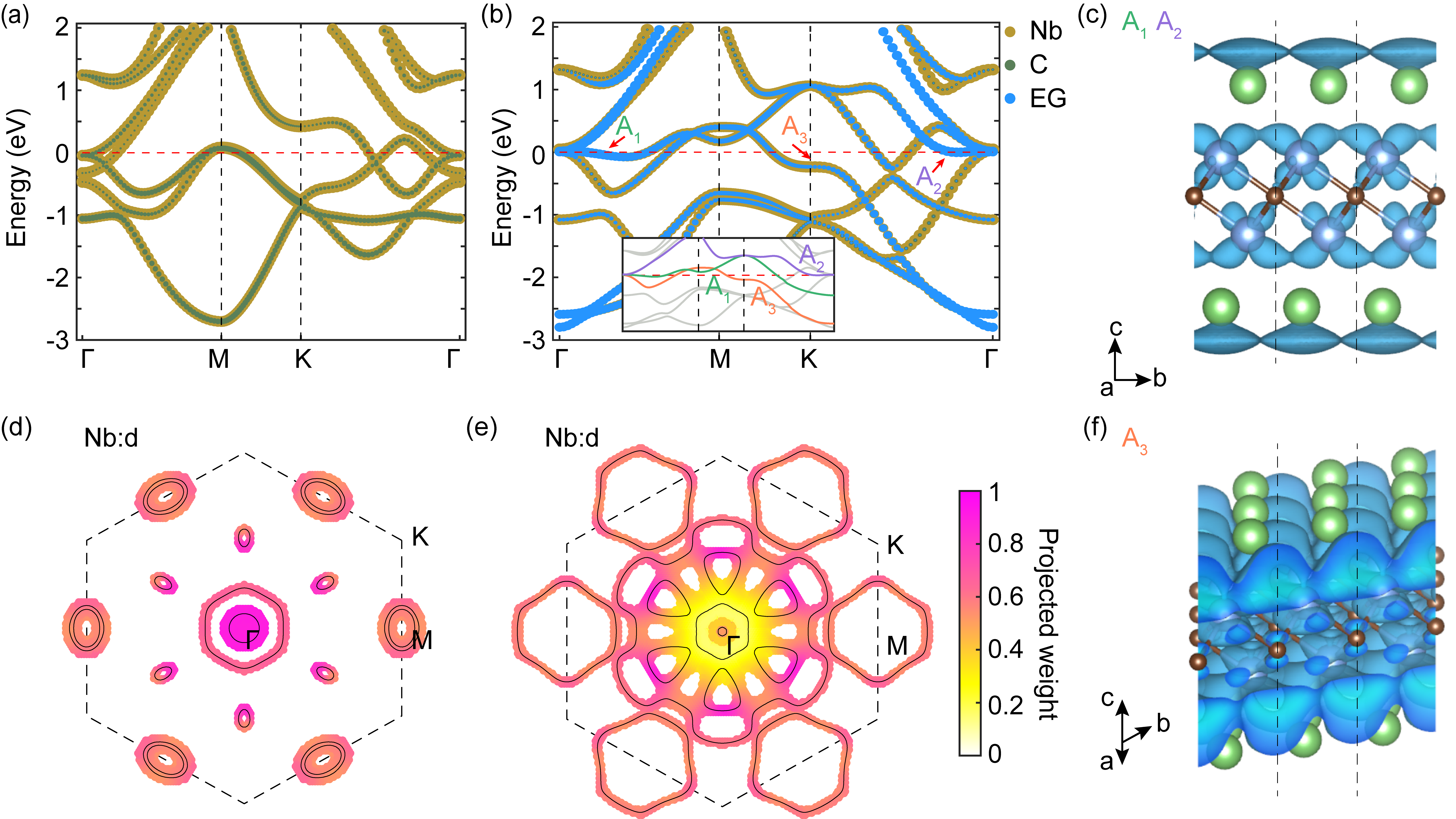}
    \caption{\label{fig2} 
    The projected band structures for monolayer $\mathrm{Nb_2C}$ (a) and $\mathrm{Nb_2CLi_2}$ (b). 
    The sizes of the colored dots are proportional to the projected weights of Nb, C, atoms and electron gas states, respectively.
    Panels (d) and (e) depict the weights of Nb-$d$ orbitals within a 50 meV energy window near $\varepsilon_{\mathrm{F}}$ for $\mathrm{Nb_2C}$ and $\mathrm{Nb_2CLi_2}$, respectively.
    The isosurface of the band decomposed charge densities distribution of selected $\boldsymbol{k}$ points on band A$_1$, A$_2$ (c) and A$_{3}$ (f).  
    The isovalues are chosen as 0.0015, and  0.0025 $e$/bohr$^3$, respectively, for better visualization.
    }
  \end{figure*}

Upon the formation of thermally and dynamically stable Nb$_2$CLi$_2$, it is important to understand the impact of Li adsorption on the electronic structure in comparison to that of the pristine monolayer.
Figs.~\ref{fig2}(a) and ~\ref{fig2}(b) display the electronic band structures of Nb$_2$C and Nb$_2$CLi$_2$, respectively, along with the projected weights onto the atomic orbitals. 
The Nb$_2$C displays a metallic band structure, with multiple bands crossing the Fermi energy ($\varepsilon_{\mathrm{F}}$). 
This results in two circular electronic pockets at the $\Gamma$ point, two elliptic hole pockets at each M point, and a  tiny elliptic electron pocket at the midpoint of each $\Gamma$--$\mathrm{K}$ path, as shown in Fig.~\ref{fig2}(d).
The Nb-$d$ orbitals primarily contribute to the electronic states around $\varepsilon_{\mathrm{F}}$, as shown in Fig.~\ref{fig2}(d).
The above results are consistent with previous studies \cite{yang2020electronic,bekaert2020first}, indicating that our computational methods are valid for the Nb$_2$C systems.

The most pronounced effect on the electronic structure after the formation of Nb$_2$CLi$_2$  is the significant increase in $N(0)$, which is evident when comparing Fig.~\ref{fig2}(d) with \ref{fig2}(e), along with the calculated density of states shown in Sec.~S1 \cite{SM}.  
This increase in $N(0)$ can be understood as follows. 
Comparing Figs.~\ref{fig2}(a) and \ref{fig2}(b) reveals that, in Nb$_2$CLi$_2$, several bands emerge near the $\varepsilon_{\mathrm{F}}$. 
Notably, two bands labeled $A_1$ and $A_2$ (see Fig.~\ref{fig2}(b) and its inset) exhibit relatively flat dispersion near $\varepsilon_{\mathrm{F}}$ along the $\Gamma\mathrm{M}$ and $\Gamma\mathrm{K}$ paths, respectively. 
This leads to the emergent electronic states around $\varepsilon_{\mathrm{F}}$ near the $\Gamma$ point, as shown in Fig.~\ref{fig2}(d). 
By analyzing the charge density distribution of the $\boldsymbol{k}$ states near $\varepsilon_{\mathrm{F}}$ from these two bands, we find they predominantly originate from Nb-$d$ orbitals and the states on the two planes outside the planes of Li atoms.
This is indicated in Figs.~\ref{fig2}(c) and ~\ref{fig2}(f), which display the isosurfaces of the charge density from the selected $\boldsymbol{k}$ points of the two bands.
Furthermore, we observed another band, labeled $A_3$, that crosses $\varepsilon_{\mathrm{F}}$ around the M points, leading to a band maximum at 0.17 eV  at each M point.
This band contributes to the hexagonal hole pockets around the M points (Fig.~\ref{fig2}(e)), which is clearly larger than the original elliptical pockets at the M points (Fig.~\ref{fig2}(d)).
A similar analysis reveals that the electronic states from this band are associated with Nb-$d$ orbitals and the interstitial states located at the centers of triangles formed by three adjacent Li atoms, as shown in Fig.~\ref{fig2}(f).
The interstitial states and the states outside the planes of Li atoms can be attributed to the low electronegativity of Li. 
We refer to them as electron gas (EG) states, since they are delocalized in certain regions of the crystal, rather than being confined around atomic cores.
Similar electronic state distributions have been observed in lithium-rich compounds \cite{zhao2023interstitial,tsuji2016structural}.
The multiple component of the Fermi surface near $\Gamma$ point is also evident from the projected Fermi surface (Fig.~\ref{fig2}(e)): the values of the projected weights related to Nb-$d$ orbitals near the $\Gamma$ point are significantly reduced compared to other wavevectors. 
Furthermore, due to electron doping induced by the adsorption of Li atoms, the two bands that dominated the elliptical hole pockets at the M points in Nb$_2$C, shift downward to become fully occupied valence bands, with their band tops located around -0.65 eV at the M points.
Therefore, the apperance of the EG states near the $\varepsilon_{\mathrm{F}}$ combined with the effect of electron doping, result in an increase in $N(0)$ from 2.91 in Nb$_2$C to 7.47 states/eV/Nb$_2$C in Nb$_2$CLi$_2$. 
It is noteworthy that the emergence of the EG states near $\varepsilon_{\mathrm{F}}$ after the introduction of Li differs from other alkali-metal-doped transition metal compounds~\cite{wu2021enhanced,zheng2020emergent,er2024emergence,habenicht2020potassium}, where the primary effect of the alkali metal atoms is electron doping, and their contribution to $N(0)$ is minimal.

\subsection{Electron-phonon coupling and superconductivity}

\begin{figure}[htb]
    \includegraphics{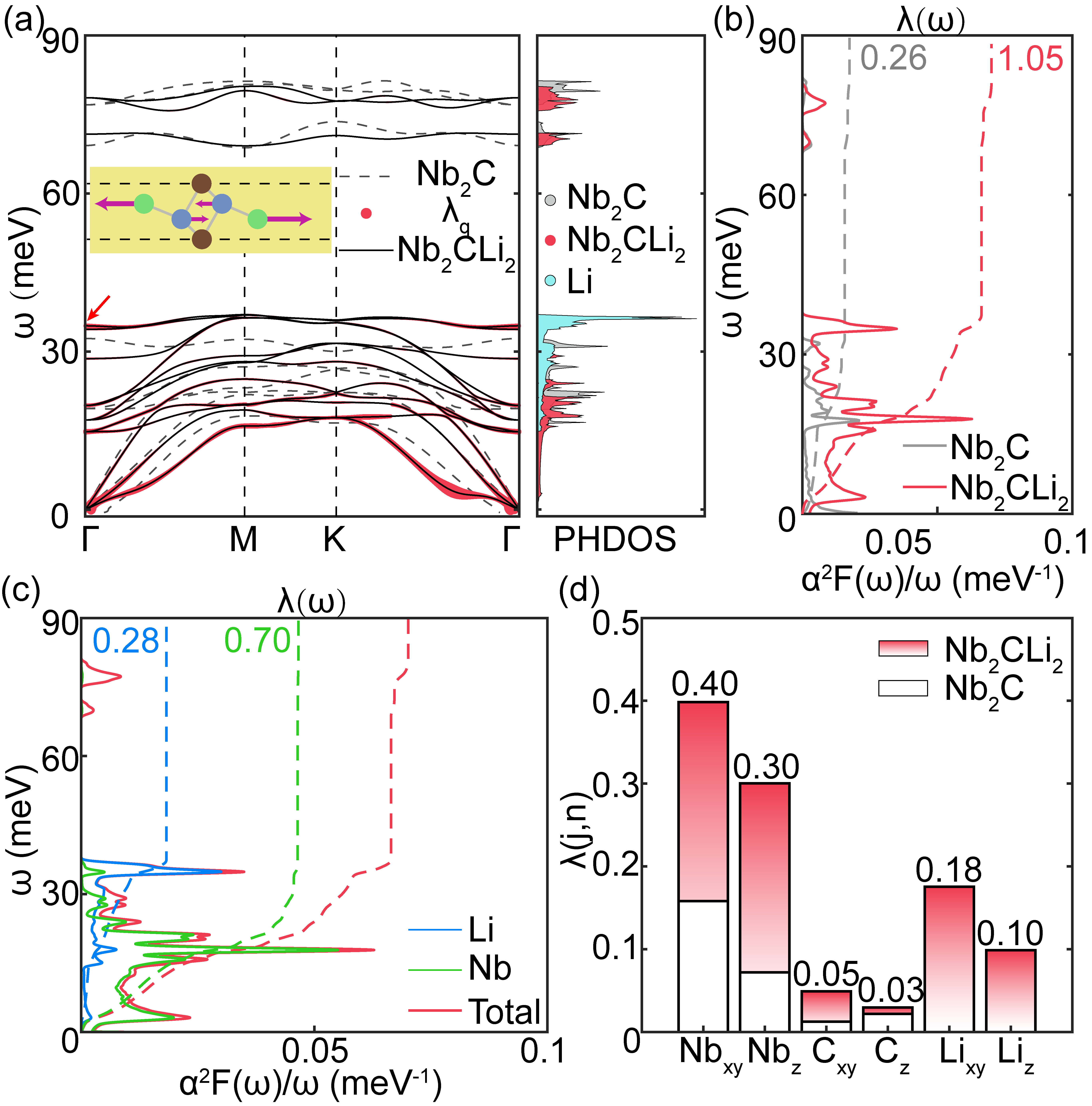}
    \caption{\label{fig3} 
    (a) Phonon dispersion for $\mathrm{Nb_2C}$ (dashed lines) and $\mathrm{Nb_2CLi_2}$ (solid lines).  
    The sizes of red dots represent the value of phonon-momentum-resolved EPC constant $\lambda_{\mathbf{q}\nu}$.
    The inset illustrates the phonon vector of the states with an energy of 35 meV at the $\Gamma$ point, indicated by a red arrow.
    The right panel shows the  PHDOS for $\mathrm{Nb_2C}$ and $\mathrm{Nb_2CLi_2}$, and the contribution from Li vibration to PHDOS, respectively. 
    (b) The $\alpha^2F(\omega)/\omega$ along with $\lambda(\omega)$ for $\mathrm{Nb_2C}$ and $\mathrm{Nb_2CLi_2}$. 
    (c)  Projected $\alpha^2F(\omega)/\omega$ onto the phonons, related to the vibrations of Nb and Li atoms in $\mathrm{Nb_2CLi_2}$. 
    (d) Atomic-vibration-resolved EPC constants for $\mathrm{Nb_2CLi_2}$, obtained by integrating the corresponding spectra in panel (c). 
    The results of $\mathrm{Nb_2C}$ are also shown for comparison.
    }
\end{figure}

Following the electronic structure analysis, we assessed the phonon dispersion and superconducting properties of Nb$_2$C.
The calculated phonon dispersion, $\omega_{\boldsymbol{q}\nu}$, as shown in Fig.~\ref{fig3}(a), demonstrates that the crystal structure is dynamically stable, as all the phonon frequencies are positive.
The calculated $\alpha^{2}F(\omega)$ and $\lambda(\omega)$ depicted in Fig.~\ref{fig3}(b) indicate that the total EPC constant $\lambda$ is only 0.26.
The corresponding isotropic $T_{\mathrm{c}}$ is calculated to be  0.01 K.
These results demonstrate the non-superconducting nature of pristine monolayer Nb$_2$C, which is consistent with recent studies \cite{kamysbayev2020covalent,yang2020electronic,zhao2014manipulation}. 

We subsequently examined the effect of Li adsorption on the lattice dynamics.
Fig.~\ref{fig3}(a) presents the $\omega_{\boldsymbol{q}\nu}$ and projected phonon density of states (PHDOS) for Nb$_2$C and Nb$_2$CLi$_2$. 
The $\omega_{\boldsymbol{q}\nu}$ of Nb$_2$C can be divided into two energy ranges: 0--33 meV, mainly arising from the heavier Nb atoms, and 68--82 meV, due to the lighter C atoms. 
After the adsorption of Li, the $\omega_{\boldsymbol{q}\nu}$  undergoes two notable changes. 
In particular, a phonon branch remains relatively flat across the entire Brillouin zone around 35 meV, which results in a peak in the PHDOS.
Meanwhile,  the phonon states of this branch display  moderate strengths of phonon momentum-dependent EPC constant ($\lambda_{\boldsymbol{q}\nu}$) in the small-momentum region close to the $\Gamma$ point. 
Visualizing the vibrations of the phonon state in this branch at the $\Gamma$ point (inset of Fig.~\ref{fig3}(a))  reveals that it is primarily associated with the out-of-plane vibration of Li atom.
This suggests that the coupled electronic states associated with these phonons primarily originate from the EG states outside the planes of Li atoms.
Another notable change is the softening of Nb phonons in the low-frequency region. 
As shown in Fig.~\ref{fig3}(a), within the energy range of 0--18 meV, the $\omega_{\boldsymbol{q}\nu}$ of Nb$_2$CLi$_2$ (solid lines) display clear redshifts compared to that of Nb$_2$C (dashed lines) in the entire Brillouin zone.
These softened phonons are with large values of $\lambda_{\boldsymbol{q}\nu}$ as indicated in Fig.~\ref{fig3}(a).
Thus, the phonon softening can be attributed to the strong EPC, which reduces the phonon energies by altering the real part of the phonon self-energy.
Similar EPC-driven phonon softening has also been observed in other systems, such as Nb$_2$CS$_2$ \cite{sevik2023superconductivity}, Li$_2$C$_6$ \cite{yang2024probing}.

According to  the above analysis, we can anticipate a significant increase in the EPC strength in Nb$_2$CLi$_2$. 
This is supported by the calculated $\alpha^{2}F(\omega)$ and $\lambda(\omega)$ as depicted in Fig.~\ref{fig3}(b). 
As shown, the total EPC constant, $\lambda$, for Nb$_2$CLi$_2$  is 1.05, exhibiting an over threefold increase compared to  Nb$_2$C (0.26).
The comparison of the calculated $\alpha^{2}F(\omega)$ with and without the constant-matrix-element approximation demonstrates that the enhancement of $\lambda$ is primarily attributed to the increase in $N(0)$, while the EPC matrix elements make a secondary contribution. 
Specifically, in the constant-matrix-element approximation for Nb$_2$CLi$_2$, the calculated value of $\lambda$, which reflects the contribution of $N(0)$, is nearly three times as large as that in Nb$_2$C. In contrast, the averaged value of the squared EPC matrix elements increases by only about 36.7\%. Further details can be found in Sec.~S6 \cite{SM}.

We further evaluate the enhancement of $\lambda$ from the perspective of atomic vibrations.
A detailed comparison of $\alpha^{2}F(\omega)/\omega$ for the two systems suggests that the EPC enhancement in Nb$_2$CLi$_2$ mainly occurs in the two energy ranges: from 0 to 26 meV, and  from 34 to 37 meV. 
In the first range, the $\lambda(\omega)$ of Nb$_2$CLi$_2$ increases by 0.78, whereas for Nb$_2$C, the increase is only 0.19.
This significant rise can be primarily attributed to the pronounced $\lambda_{\boldsymbol{q}\nu}$ associated with the Nb phonons, as previously mentioned. 
Further confirmation comes from the calculated atomic-vibration-resolved Eliashberg function $\alpha^{2}F(\omega,j,\hat{n})$, which is obtained by projecting $\alpha^{2}F(\omega)$ onto the $j$th atom in the vibrational direction of $\hat{n}$. 
As shown in Fig.~\ref{fig3}(c), the spectrum of $\alpha^{2}F(\omega,\mathrm{Nb})/\omega$ closely aligns with the $\alpha^{2}F(\omega)/\omega$ in this energy range,  indicating that Nb vibrations dominate the contribution to $\lambda(\omega)$.
In the second range, the increase in the EPC constant for Nb$_2$CLi$_2$ is 0.13, which is mainly associated with the Li vibrations based on similar analysis.
In contrast, the increase for Nb$_2$C is negligible.
Integrating $\alpha^{2}F(\omega,j,\hat{n})/\omega$ over $\omega$ leads to the atomic-vibration-resolved EPC constant $\lambda(j,\hat{n})$, which quantifies the EPC contributions from the vibrations of the $j$-th atom in the $\hat{n}$ direction. 
As illustrated in Figs.~\ref{fig3}(d) and~\ref{fig3}(c), the contributions to $\lambda$ arising from  Nb and Li vibrations are 0.70 and 0.28, respectively, comprising 66.7\% and 26.7\% of the total $\lambda$. 
The above discussion indicates that the significant enhancement in EPC of Nb$_2$CLi$_2$ primarily arises from the contributions of low-energy phonons associated with Nb vibrations, with Li vibrations making a secondary contribution.

\begin{figure}[h]
    \includegraphics{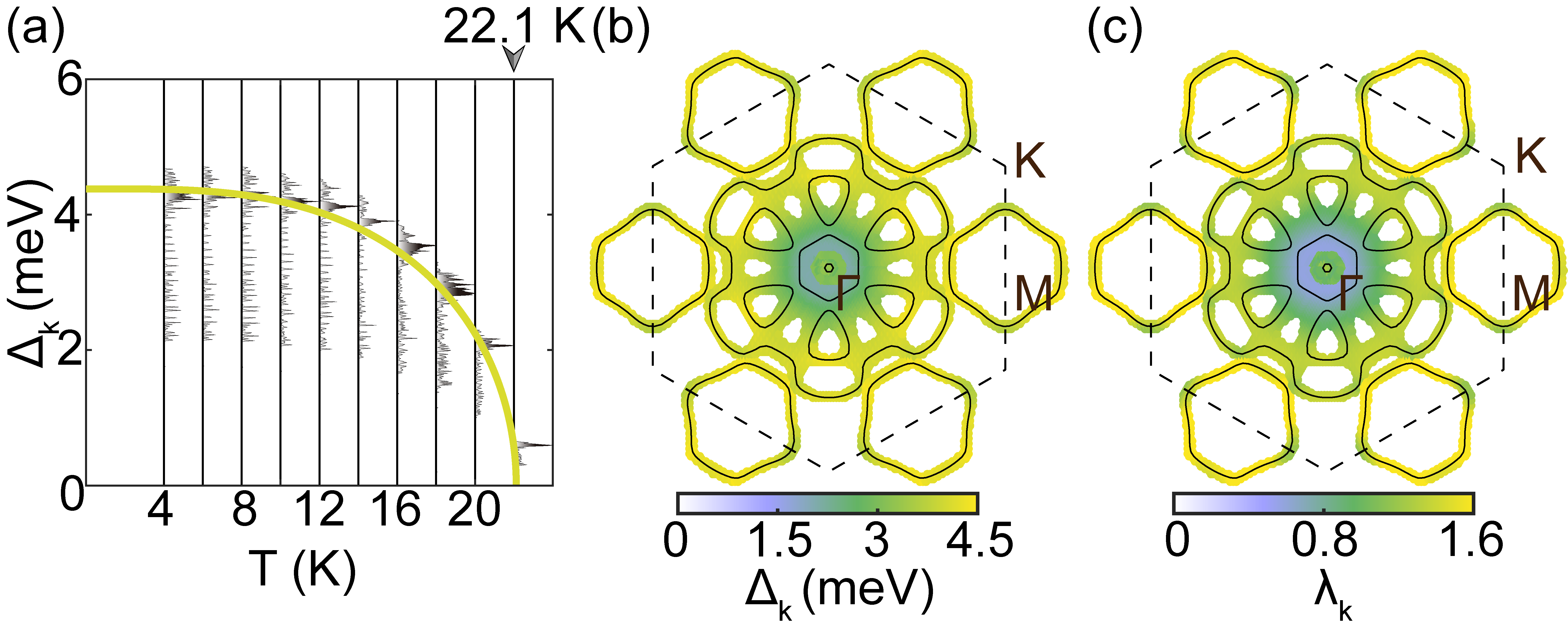}
    \caption{
    \label{fig4} 
    (a) Histograms of $\Delta(\boldsymbol{k},T)$ for $\mathrm{Nb_2CLi_2}$. 
    The yellow curve is the BCS fit. 
    (b) The distribution of the $\Delta_{\boldsymbol{k}}$ at $T = 4$ K in an extended BZ.
    Only the $\boldsymbol{k}$ states located within an energy window of 50 meV near Fermi level are shown. 
    Solid lines represent the Fermi surface. 
    (c) A similar plot to panel (b), illustrating the distribution of electronic momentum-resolved EPC constants, $\lambda_{\boldsymbol{k}}$.   
    }
\end{figure}
The significant increase of EPC suggests a potential enhancement in the superconducting properties of Nb$_2$CLi$_2$.
Using the full Allen-Dynes formula \cite{allen1975transition}, the superconducting $T_{\mathrm{c}}$ of Nb$_2$CLi$_2$ was calculated to be 15.9 K (refer to Sec.~S3 \cite{SM} for details). 
Given that low-dimensional systems with anisotropic Fermi surfaces typically exhibit anisotropic EPC, which results in anisotropic superconductivity, we further explored its anisotropic superconducting properties. 
This included analyzing the electronic momentum-resolved EPC constants ($\lambda_{\boldsymbol{k}}$) and the superconducting gap ($\Delta_{\boldsymbol{k}}$) by solving the anisotropic Migdal-Eliashberg equations at various temperatures \cite{ponce2016epw,mostofi2008wannier90}.
As depicted in Fig.~\ref{fig4}(a), the values of $\Delta_{\boldsymbol{k}}$ decrease with increasing temperature, vanishing at 22.1 K, which indicates that the anisotropic $T_{\mathrm{c}}$ is approximately 22.1 K. 
A more detailed analysis of the $\Delta_{\boldsymbol{k}}$ structure reveals a single-gap characteristic, but with an extended energy distribution below the $T_{\mathrm{c}}$. 
For instance, at $T=4$ K, the values of $\Delta_{\boldsymbol{k}}$ range from 2.1--4.7 meV, resulting in an energy span of 2.6 meV. 
This highlights strong anisotropic superconductivity, which is consistent with a previous study in which anisotropic superconductivity was observed in functional-group-modified Nb$_2$C \cite{xu2024anisotropic}. 
Further analysis suggests that the energy range can be divided into two regions: (I) a high-energy region from 4 to 4.7 meV, where a prominent peak appears, indicating a significant concentration of $\Delta_{\boldsymbol{k}}$ values; and (II) a low-energy region from 2 to 4 meV, where $\Delta_{\boldsymbol{k}}$ is uniformly distributed.
By visualizing the distribution of the $\Delta_{\boldsymbol{k}}$  across the Brillouin zone, as shown in Fig.~\ref{fig4}(b), we find that $\Delta_{\boldsymbol{k}}$ in the high-energy section mainly originate from Fermi pockets at the M points, as well as from the outermost Fermi pockets at $\Gamma$ point. 
A comparison with Fig.~\ref{fig2}(e) indicates that these electronic states are primarily derived from Nb-$d$ orbitals. 
A similar analysis shows that the $\Delta_{\boldsymbol{k}}$ in the low-energy region are associated with the electronic states near the Fermi pockets close to the $\Gamma$ point. 
These states are related to the Nb-$d$ orbitals and the EG states outside the planes of Li atoms, as analyzed previously.
The calculated $\lambda_{\boldsymbol{k}}$, depicted in Fig.~\ref{fig4}(c), also shows consistent anisotropic characteristics. 
Therefore, we have shown that after the adsorption of Li atoms, the system transforms from a conventional metal into a single-gap superconductor with extended distribution of superconducting gap values, achieving a $T_{\mathrm{c}}$ of 22.1 K. 
The strong anisotropic superconductivity  is due to the electronic states from  Nb-$d$ orbitals, which exhibit relatively large $\Delta_{\boldsymbol{k}}$, and the EG states, which contribute to relatively small  $\Delta_{\boldsymbol{k}}$.

\subsection{The effect of strain}

\begin{figure}[h]
  \includegraphics{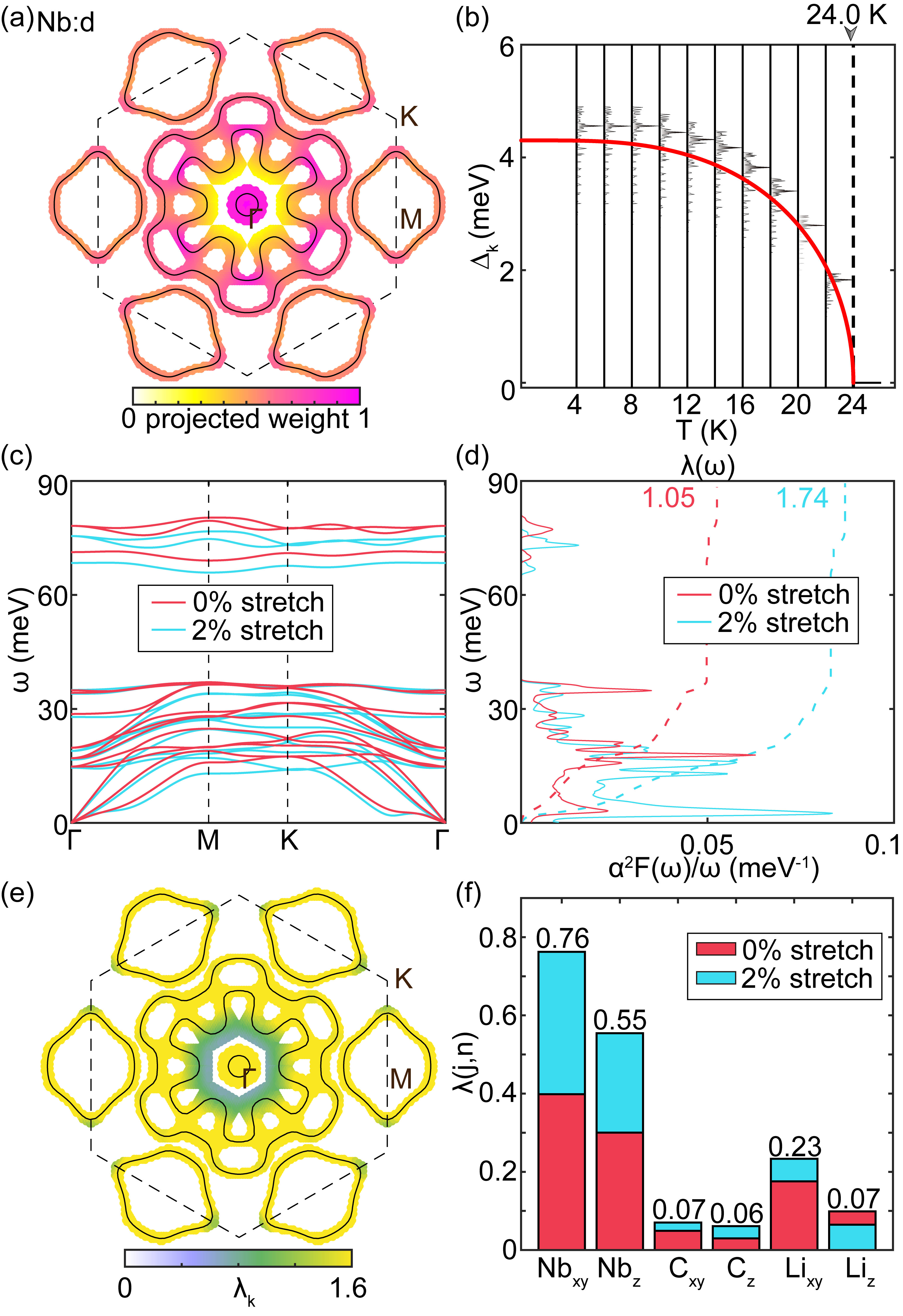}
  \caption{\label{fig5} 
  (a) Projected Fermi surface onto the Nb-$d$ orbitals for 2\%-stretched $\mathrm{Nb_2CLi_2}$ within an energy window of 50 meV near Fermi level.  
  (b) Histograms of $\Delta (\boldsymbol{k},T)$ for 2\%-stretched $\mathrm{Nb_2CLi_2}$. 
  The Red curve is the a BCS fit of the gap.
  The phonon dispersion (c) and $\alpha^2F(\omega)/\omega$ (d) for strain-free and 2\%-stretched $\mathrm{Nb_2CLi_2}$. 
  (e) The distribution of Electronic-momentum-resolved EPC constants ($\lambda_{\boldsymbol{k}}$) around Fermi surface in an extended Brillouin zone, for $\mathrm{Nb_2CLi_2}$.
  Only the $\boldsymbol{k}$ states within an energy of $\pm$50 meV related to the  Fermi level are shown. 
  (f) Atomic-vibration-resolved EPC constants for strain-free and 2\%-stretched $\mathrm{Nb_2CLi_2}$.
  }
\end{figure}

Considering that the in-plane vibrations of Nb atoms in Nb$_2$CLi$_2$ contribute significantly to the EPC (Fig.~\ref{fig3}(d)), and that thin-layer materials are susceptible to strain due to external environments, we further investigated the effect of biaxial strain. 
Sec.~S3 \cite{SM} demonstrates the variation of the EPC constant and McMillan $T_{\mathrm{c}}$ with strains. 
Both quantities exhibit similar trends: remaining nearly constant under compressive strain but increasing rapidly under tensile strains.
At 2\% tensile strain, the EPC constant and $T_{\mathrm{c}}$ reach 1.74 and 22.4 K, respectively. 
Further stretching leads to dynamical instability of the structure. 
Thus, Nb$_2$CLi$_2$ exhibits optimal superconductivity under 2\% in-plane tensile strain, which we will refer to as 2\%-stretched Nb$_2$CLi$_2$.
As shown in Fig.~\ref{fig5}(b), further anisotropic EPC calculation  reveals that the variation in distributions of  $\Delta_{\boldsymbol{k}}$ with temperature indicates an anisotropic $T_{\mathrm{c}}$ of approximately 24 K, which is higher than that of strain-free Nb$_2$CLi$_2$. 
This $T_{\mathrm{c}}$ is significantly higher than most functional Nb$_2$C such as Nb$_2$CS$_2$ (6.4 K) \cite{kamysbayev2020covalent}, Nb$_2$CCl$_2$ (5.2 K)\cite{wang2022role}, Nb$_2$CH$_2$ (5.0 K) \cite{bekaert2022enhancing}, and lower than doped monolayer Nb$_2$CCl$_2$  \cite{sevik2023superconductivity} and Nb$_2$CSO  \cite{zhang2023effect}.

The increase in the $T_{\mathrm{c}}$ is mainly due to the enhanced EPC strength driven by the changes in the electronic structure around the $\Gamma$ point,  and the softening of low-frequency phonons as discussed below. 
In comparing Figs.~\ref{fig5}(a) and \ref{fig2}(e), there is a clear decrease in the contribution from EG states and an increased contribution from the Nb-$d$ states near the $\Gamma$ point after applying strain.
This is corroborated by the analysis of electronic band structure changes before and after strain application shown in Sec.~S2 \cite{SM}.
Previous analysis of superconducting properties in strain-free Nb$_2$CLi$_2$ showed that Nb-$d$ states exhibit higher  $\lambda_{\boldsymbol{k}}$ compared to the EG states. 
Thus, the increased contribution from Nb-$d$ states is expected to enhance the overall EPC strength.
This is evidenced by comparing the electronic momentum-resolved EPC constants ($\lambda_{\boldsymbol{k}}$) before and after stretching (Figs.~\ref{fig5}(e) and \ref{fig4}(c)), which show noticeable enhanced EPC strength around the $\Gamma$ point in 2\%-stretched Nb$_2$CLi$_2$.
Furthermore, comparing the phonon spectra before and after the strain (Fig.~\ref{fig5}(c)) reveals significant softening of low-frequency phonons in 2\%-stretched Nb$_2$CLi$_2$ along the entire Brillouin zone.
The softening of phonons can be attributed to their enhanced coupling with electronic states, as shown in Sec.~S4 \cite{SM}. 
Fig.~S6 \cite{SM} further demonstrates that these softening phonons are primarily contributed by Nb vibrations.
The softening of these phonons, combined with the increased contribution from Nb-$d$ orbitals near the Fermi surface, raises the $\lambda$ of 2\%-stretched Nb$_2$CLi$_2$ from 1.05 in strain-free states to 1.74. 
Furthermore, we noticed the Fermi surface anisotropy in 2\%-stretched Nb$_2$CLi$_2$  is suppressed compared to its strain-free state, owing to the reduced contribution from the EG states.  
This leads to changes in the superconducting gap structure and a closer alignment between the McMillan and anisotropic $T_{\mathrm{c}}$ values, which are 22.4 and 24 K, respectively.
A careful examination of the energy distribution of $\Delta_{\boldsymbol{k}}$  in Fig.~\ref{fig5}(b) reveals a more concentrated distribution compared to that of its strain-free state (Fig.~\ref{fig4}(a)).
For example, at a temperature of 4 K, most $\Delta_{\boldsymbol{k}}$ values are concentrated around 4.7 meV, with an energy window broadening of approximately 2.1 meV (Fig.~\ref{fig5}(b)), which is smaller than the 2.6 meV in the strain-free  Nb$_2$CLi$_2$. 
This suggests that the anisotropy of the superconductivity in 2\%-stretched Nb$_2$CLi$_2$ diminishes as the anisotropy of the Fermi surface decreases.
The above results lead us to conclude that the enhanced in $T_{\mathrm{c}}$ in 2\%-stretched Nb$_2$CLi$_2$ is due to the enhanced contribution of Nb-$d$ orbitals to the Fermi surface with reduced Fermi surface anisotropy, and the red shift of the Nb-vibration-derived phonon states in the low-energy range.

\section{Conclusion}

In summary, we have computationally studied the crystal structure, electronic structure, phonon, EPC, and superconductivity in monolayer Nb$_2$CLi$_2$. 
The computed formation energies and convex hull indicate that for Nb$_2$C, the adsorbed Li atoms are energetically favorable to occupy the sites overhead the Nb atoms on their opposite side. 
This increases the contribution of Nb-$d$ orbitals to the Fermi surface due to the electron doping, and leads to the emergence of  EG states near the Fermi surface, arising from the low electronegativity of Li.
These combined effects significantly enhance the $N$(0) and $\lambda$ in Nb$_2$CLi$_2$,  giving rise to a metal-to-superconductor transition with an anisotropic $T_{\mathrm{c}}$ of 22.1 K.
The calculated superconducting gap structure suggests that Nb$_2$CLi$_2$ is a single-gap superconductor with a significantly anisotropic gap distribution. 
The Nb-$d$ orbitals contribute most prominently to the superconductivity, while the EG states provide a secondary contribution.
Applying a 2\% biaxial tensile strain leads to the increase of $T_{\mathrm{c}}$ up to 24.0 K and the reduced anisotropy, due to the increasing contribution from the Nb-$d$ orbitals and the softening of Nb-vibration-derived low-frequency phonons. 
This $T_{\mathrm{c}}$ is relatively high among the decorated Nb$_2$C systems. 
Considering the experimental feasibility of Nb$_2$C and the fact that previous related works primarily focus on the decoration of halogen and chalcogen atoms, as well as the experimental realization of alkaline-metal atom depositions \cite{ludbrook2015evidence,kondekar2017situ,li2020reduced,yang2022high,lee2024interaction}, our study enhances the understanding of the superconducting properties in monolayer Nb$_2$C and offers a new, experimentally feasible approach to achieve a significantly higher $T_{\mathrm{c}}$.

\begin{acknowledgements} 
    This work is supported by National Natural Science Foundation of China 11804118, Guangdong Basic and Applied Basic Research Foundation (Grant No. 2025A1515010219, 2021A1515010041), and the Science and Technology Planning Project of Guangzhou (Grant No. 202201010222). 
    The Calculations were performed on  high-performance computation cluster of Jinan University, and Tianhe Supercomputer System.
  \end{acknowledgements}

  \bibliographystyle{apsrev4-2}
  %\bibliography{main} 
  
%

% \textcolor{red}{this}

\end{document}